\documentclass[]{spie}  

\usepackage{amsmath,amsfonts,amssymb}
\usepackage{graphicx}
\usepackage[colorlinks=true, allcolors=blue]{hyperref}

\usepackage{gensymb}

\title{Revisiting the Borde-Traub focal plane wavefront estimation technique for exoplanet direct imaging}

\author[a,b,*]{Axel Potier}
\author[a]{A J Eldorado Riggs}
\author[a]{Garreth Ruane}
\author[a]{Phillip K. Poon}
\author[a]{Matthew Noyes}
\author[a]{Greg W. Allan }
\author[a]{Alexander Walter}
\author[a]{Camilo Mejia Prada}
\author[c]{Raphael Galicher}
\author[c]{Johan Mazoyer}
\author[c]{Pierre Baudoz}

\affil[a]{Jet Propulsion Laboratory, California Institute of Technology, 4800 Oak Grove Drive, Pasadena, CA 91109}
\affil[b]{Space Research \& Planetary Sciences Division, Physics Institute, University of Bern, Sidlerstrasse 5, 3012 Bern, Switzerland}
\affil[c]{LESIA, Observatoire de Paris, Université PSL, CNRS, Université Paris Cité, Sorbonne
Université, 5 place Jules Janssen, 92195 Meudon, France}

\authorinfo{*Send correspondence to axel.potier@unibe.ch}

\begin{document} 
\maketitle

\begin{abstract}
Direct imaging of exoplanets relies on complex wavefront sensing and control architectures. In addition to fast adaptive optics systems, most of the future high-contrast imaging instruments will soon be equipped with focal plane wavefront sensing algorithms. These techniques use the science detector to estimate the static and quasi-static aberrations induced by optical manufacturing defects and system thermal variations. Pair-wise probing (PWP) has been the most widely used, especially for space-based application and will be tested at contrast levels of $\sim$1e-9 on-sky along with the future coronagraph instrument onboarding the Roman Space Telescope. This algorithm leans on phase diversities applied on the deformable mirror that are recorded in pairs. A minimum of two pairs of probes are required per bandwidth. An additional unprobed image is also recorded to verify the convergence rate of the correction. Before PWP, Borde \& Traub proposed a similar algorithm that takes advantage of the unprobed image in the estimation process to get rid of the pair diversity requirement. In this work, we theoretically show that this latter technique should be more efficient than PWP when the convergence time is not limited by photon noise. We then present its performance and practical limitations on coronagraphic testbeds at JPL and exhibit a first on-sky control of non-common path aberrations with such method on VLT/SPHERE.
\end{abstract}

\keywords{Coronagraphy, High-contrast imaging, Wavefront sensing and control, Exoplanets}


\section{Introduction}
Direct imaging of exoplanets remains a major challenge in modern astrophysics because of the high contrast and the close vicinity of a star and its planetary companions. On ground-based facilities, coronagraphs are used in association with state-of-the-art adaptive optics systems to attenuate the starlight and reveal the presence of their faint escorts in images\cite{Galicher2023}. In the near future, such high-order wavefront sensing and control (HOWS\&C) capabilities will also equip the coronagraph instrument of the Roman Space Telescope (RST), which serves as a technological demonstrator for the upcoming Habitable World Observatory (HWO). 

Before any science acquisitions, a long optical correction procedure will be required. The science detector will be used as a wavefront sensing camera to estimate the optical aberrations at the coronagraph plane\cite{Malbet1995}. A few focal plane wavefront sensors (FPWS) have been proposed. Taking advantage of a Young's double slit interferometer built in the coronagraph, the Self-Coherent Camera spatially modulates the starlight to estimate the electric field (E-field)\cite{Baudoz2006}~. For RST, the E-field will be estimated through a temporal modulation of the starlight \cite{Kasdin2020} performed by applying known voltages to the deformable mirrors (DM), whose function is therefore not limited to wavefront correction. Such method, that uses pairs of probes set on the DM is called Pair-Wise probing (PWP) \cite{GiveOn2007SPIE}. In association with the electric field conjugation (EFC) correction algorithm, PWP has already demonstrated contrast levels (the ratio between the companion flux and its host star) of a few $10^{-10}\,$ \cite{Seo2019} that is relevant to the imaging of exoEarths.

Because it interrupts the science acquisition, the HOWS\&C loop could limit the mission yield. Indeed, PWP + EFC usually requires some dozens of iterations to obtain deep contrast values. During each estimation step, at least 6 images per spectral sub-bands must be taken to create a correction in the full field of view (or Dark Hole, DH), with 5 sub-bands usually defined to split a 20\% spectral bandwidth.
Additionally, PWP needs longer and longer exposure times through the correction to sense smaller errors beyond the stellar shot noise. For RST, each iteration should take up to 5 hours of telescope time\cite{Zhou2020}.

Before PWP, Bord\'e \& Traub\cite{Borde2006} proposed a similar algorithm where probe images are recorded individually rather than in pairs, decreasing the number of images required per iteration. This work aims to further study this technique through a deep comparison with the well-known PWP algorithm, especially in term of efficiency and convergence time, for the sake of increasing the duty cycle in operations. In Sec.~\ref{subsec:method}, we describe the Bord\'e \& Traub probing (BTP) algorithm and compare PWP and BTP sensitivities in Sec.~\ref{subsec:sensitivity}. We also discuss the impact of model errors in Sec.~\ref{subsec:modelerror} and propose a model free version of BTP in Sec.~\ref{subsec:implicit}. We then perform BTP in stabilized environment for various coronagraphs as explained in Sec.~\ref{sec:JPLexperiment}. Because we think BTP is more promising in ground-based applications, we finally present the first on-sky correction of BTP with VLT/SPHERE in Sec.~\ref{sec:SPHEREexperiment}.

\section{BTP algorithm theory}
\label{sec:Theory}
\subsection{Method}
\label{subsec:method}
\subsubsection{General formalism}
In this section, we aim to explain the BTP algorithm. Letting $E_k$ be the electric field in the focal plane induced by the $k^{\text{th}}$ probe applied on the DM $\Psi_k$, $E_{ab}$ being the stellar E-field to be estimated and A the entrance pupil, we can write
\begin{equation}
    E_k = E_{ab} + iC\{A\Psi_k\}, 
\end{equation}
through the assumption of small phase perturbation applied on the DM\cite{Borde2006,GiveOn2007}~. Here, $C\{\}$ is the linear operator representing the propagation through the coronagraph instrument from the pupil to the focal plane. The intensity of such probe $I_k$ is therefore the simple squared modulus of the E-field.

\subsubsection{Pair-Wise Probing}
For PWP, we take the difference of a positive $I_{k}^+$ and a negative probe $I_{k}^-$ to conserve the cross term that is linear to the speckle E-field:
\begin{equation}
    \Delta I_k = I_{k}^+ - I_{k}^- = 4\Re(E_{ab}^*iC\{A\Psi_k\}) , 
\end{equation}
where $\Re$ stands for the real part of the complex number. Finally, by granting a full set of $n_p$ pairs of probes, we can write
\begin{equation}
\label{eq:PWP}
\begin{pmatrix}
\Delta I_1 \\
... \\
\Delta I_{n_p}
\end{pmatrix} = 4 \begin{pmatrix}
\Re(| iC\{A\Psi_1\} |) & \Im(| iC\{A\Psi_1\} |) \\
... \\
\Re(| iC\{A\Psi_{n_p}\} |) & \Im(| iC\{A\Psi_{n_p}\} |)
\end{pmatrix}\begin{pmatrix}
E_R \\
E_I
\end{pmatrix},
\end{equation}
where $E_R$ and $E_I$ are the real and imaginary parts of the stellar E-field.

This equation can be inverted to retrieve the complex stellar E-field. Applying the formalism used in previous works\cite{Groff2015}, we can write:
\begin{equation}
\label{eq:inversion}
    x = (H^TH^T)^{-1}H^Tz.
\end{equation}
where:
\begin{equation}
    x = \begin{pmatrix}
E_R \\
E_I  
    \end{pmatrix},
\end{equation}

\begin{equation}
\label{eq:model}
    H = 4 \begin{pmatrix}
\Re(| iC\{A\Psi_1\} |) & \Im(| iC\{A\Psi_1\} |) \\
... \\
\Re(| iC\{A\Psi_{n_p}\} |) & \Im(| iC\{A\Psi_{n_p}\} |)
\end{pmatrix}.
\end{equation}
and
\begin{equation}
\label{eq:PWPmeas}
    z^{PWP} = \begin{pmatrix}
\Delta I_1 \\
... \\
\Delta I_{n_p}
\end{pmatrix}.
\end{equation}
Usually $H$ is obtained through a propagation model of the probes.

\subsubsection{Bord\'e \& Traub Probing}
BTP takes advantage of the unprobed speckle image $I^0 = | E_{ab} |^2$ to derive a similar equation. By subtracting one probe image with the unprobe image, we find:
\begin{equation}
    I_k^+ - I^0 = | iC\{A\Psi_k\} |^2 + 2\Re{E_{ab}^*iC\{A\Psi_k\}}, 
\end{equation}
Then, again expanding to a full set of $n_p$ probes, we retrieve a similar expression as in PWP (eq.~\ref{eq:PWP}):
\begin{equation}
\label{eq:BTP}
\begin{pmatrix}
I_1^+ - I^0 - | iC\{A\Psi_1\} |^2 \\
... \\
I_{n_p}^+ - I^0 - | iC\{A\Psi_{n_p}\} |^2
\end{pmatrix} = 2 \begin{pmatrix}
\Re(| iC\{A\Psi_1\} |) & \Im(| iC\{A\Psi_1\} |) \\
... \\
\Re(| iC\{A\Psi_{n_p}\} |) & \Im(| iC\{A\Psi_{n_p}\} |)
\end{pmatrix}\begin{pmatrix}
E_R \\
E_I
\end{pmatrix}.
\end{equation}
This equation can be inverted with the same formalism as in eq.~\ref{eq:inversion} where $x$ and $H$ are equivalent for both PWP and BTP. Here the difference between BTP and PWP only comes from the last term:
\begin{equation}
\label{eq:BTPmeas}
    z^{BTP} = 2 \begin{pmatrix}
I_1^+ - I^0 - | iC\{A\Psi_1\} |^2 \\
... \\
I_{n_p}^+ - I^0 - | iC\{A\Psi_{n_p}\} |^2
\end{pmatrix}.
\end{equation}

Equation \ref{eq:PWPmeas} and \ref{eq:BTPmeas} show that PWP uses pairs of probes to estimate the E-field while BTP requires one probe image as well as the unprobed E-field $I^0$, leading to $n_p-1$ less images per subband to obtain an estimate. Additionally, $I^0$ is usually measured at each PWP+EFC iteration to assess proper convergence of the algorithm. It can therefore be reused for free to estimate the E-field thanks to BTP and save $n_p$ images per subband per iteration. Equation~\ref{eq:BTPmeas} also introduces $| iC\{A\Psi_{n_p}\} |^2$. Its value can either be numerically modeled as in eq.~\ref{eq:model} or could be measured thanks to the following expression:
\begin{equation}
\label{eq:modelfreeprobe}
   | iC\{A\Psi_{k}\} |^2 =  \frac{I_{k}^+ - I_{k}^-}{2} - I^0
\end{equation}
requiring an additional negative probe image (as in PWP) each time the probes are modified during convergence. This model free technique would therefore be equivalent to PWP if the probes are modified at every single iteration.

\subsection{Sensitivity comparison with Pair Wise Probing}
\label{subsec:sensitivity}
In this section, we assess the sensitivity of BTP to stellar shot noise following the formalism used in previous studies\cite{Groff2015}~. We compare it to PWP and derive the total exposure time required to reach a similar contrast for both the algorithms. 

For each pixel on the corrected region, we first choose a set of probes that deliver an equal amplitude $p = |iC\{A\Psi\}|$. The variance $\sigma^2\{\}$ of the real and imaginary part of the E-field can be written as\cite{Groff2015}~:
\begin{equation}
\label{eq:totalvariance}
\begin{pmatrix}
\sigma^2\{E_R\} \\
\sigma^2\{E_I\}
\end{pmatrix} = \left(\frac{1}{4p}\right)^2\left(\frac{2}{n_p}\right)\sigma^2\{\Delta I\}\begin{pmatrix}
1 \\
1
\end{pmatrix} + \frac{\sigma^2\{p\}}{p^2}\begin{pmatrix}
E_R^2 \\
E_I^2
\end{pmatrix},
\end{equation}
where, in the case of BTP, $\Delta I$ can be written as:
\begin{equation}
\label{eq:deltaI_glob}
    \Delta I = 2(I^+ - I^0 - | iC\{A\Psi\} |^2).
\end{equation}
Let first assume that $\sigma^2\{p\}<<p^2$ i.e. the model uncertainty on the probes is small enough for the first term in eq.~\ref{eq:totalvariance} to dominate:
\begin{equation}
\label{eq:totalvariance2}
    \sigma^2\{E_R\} = \sigma^2\{E_I\} = \left(\frac{1}{4p}\right)^2\left(\frac{2}{n_p}\right)\sigma^2\{\Delta I\},
\end{equation}

These terms lead to the contrast limitation $N_C$ due to the diverse error sources:
\begin{equation}
\label{eq:NC}
    N_C = (\sigma^2\{E_R\} + \sigma^2\{E_I\})/I_{pk},
\end{equation}
where $I_{pk}$ is the acquired number of photoelectrons at the peak of the non-coronagraphic point spread function of a star whose flux is $F_{pk}$ and during a given exposure time $t$:
\begin{equation}
    I_{pk} = F_{pk}t.
\end{equation}
It has been proven that large probes outperform small probes in the context of PWP\cite{Groff2015}. In the following, we therefore assume large probes to derive a simpler expression from eq.~\ref{eq:totalvariance2}. With $|p|>>|E_{ab}|$, we can write that $I^+\sim p^2$ whose unit is photoelectrons. Also, according to Poisson statistics, the variance of the distribution is equal to its expected value: $\sigma^2\{I^+\} \sim p^2 $ whose unit is now in photoelectrons$^2$. These expressions are used to derive $\sigma^2\{\Delta~I\}$ in the cases that follow.

\subsubsection{If \texorpdfstring{$| iC\{A\Psi\} |^2$}{Lg} is computed from the propagation model}
\label{subsubsec:modelbased}
We first assume that $| iC\{A\Psi\} |^2$ is modeled and does not require any additional image. Then, according to eq.~\ref{eq:deltaI_glob}:
\begin{equation}
\label{eq:eq17}
    \sigma^2\{\Delta I\} = 4\Big(\sigma^2\{I^+\} + \sigma^2\{I^0\} + \sigma^2\{p^2\} \Big)
\end{equation}
We assume $p$ is normally distributed: $p \sim \mathcal{N}(p,\,\sigma^{2}(p))$. Then, according to the non-central moments of p, we can write the variance of $p^2$ as:
\begin{equation}
    \sigma^2\{p^2\} = 4p^2\sigma^2\{p\} + 2\sigma^4\{p\}.
\end{equation}
However, we have already assumed that $\sigma^2\{p\}<<p^2$ such that $\sigma^2\{p^2\} \sim 4p^2\sigma^2\{p\}$. We can therefore write eq.~\ref{eq:eq17} as:
\begin{equation}
    \sigma^2\{\Delta I\} = 4p^2\Big(1+4\sigma^2\{p\}\Big)
\end{equation}
Then, according to eq.~\ref{eq:totalvariance2}:
\begin{equation}
\label{eq:varianceBTP}
    \sigma^2\{E_R\} = \frac{1}{2n_p}(1+4\sigma^2\{p\}).
\end{equation}
Again, first assuming the error on the probe model to be small ($\sigma^2\{p\}<<1$), we can use this expression to derive eq.~\ref{eq:NC}:
\begin{equation}
    N_C = \frac{1}{n_pF_{pk}t} = \frac{1}{F_{pk}t_{cal}},
\end{equation}
where we simply use the fact that the total exposure time applied to the calibration $t_{cal}$ is the total number of acquired images multiplied by the individual exposure time $t_{exp}$. In this case, since the number of images equals the number of probes, we have: $t_{cal} = n_pt_{exp}$. Therefore the total exposure time per iteration required to obtain a contrast $N_C$ nearby a star whose flux is $F_{pk}$ is the sum of the exposure time needed by the algorithm and the total overhead time $t_{toh}$:
\begin{equation}
\label{eq:timeBTP}
    t_{tot} = t_{cal} + t_{toh} = \frac{1}{F_{pk}N_C} + n_pt_{oh},
\end{equation}
where $t_{oh}$ is the overhead time to take one image. This overhead time can be caused either by the detector read out time, a slow DM transfer function, a limited bandwidth to transfer images to the real time calculator (for space-based applications), or the time required to freeze the atmospheric turbulence (for ground-based applications\cite{Singh2019}~).

\subsubsection{If \texorpdfstring{$| iC\{A\Psi\} |^2$}{Lg} is measured at each iteration}
Now, we assume that $| iC\{A\Psi\} |^2$ is measured through eq.~\ref{eq:modelfreeprobe} at each iteration, requiring the acquisition of the generic negative probe image $I^-$. Then, according to eq.~\ref{eq:deltaI_glob}:
\begin{equation}
    \sigma^2\{\Delta I\} = \sigma^2\{2(I^+-I^0-\frac{I^+-I^-}{2}+I^0)\}=\sigma^2\{I^+ - I^-\}=2p^2
\end{equation}
Then, according to eq.~\ref{eq:totalvariance2}:
\begin{equation}
    \sigma^2\{E_R\} = \frac{1}{4n_p}
\end{equation}
Again, we can now compute eq.~\ref{eq:NC}, reminding the reader that in this case, pairs of probes are required at each iterations hence $t_{cal} = 2n_pt_{exp}$:
\begin{equation}
    N_C = \frac{1}{2n_pF_{pk}t_{exp}} = \frac{1}{F_{pk}t_{cal}}.
\end{equation}
As noticed before, this expression is exactly equivalent to PWP's sensitivity\cite{Groff2015}. We can now compute the total exposure time:
\begin{equation}
\label{eq:timePWP}
    t_{tot} = t_{cal} + t_{toh} = \frac{1}{F_{pk}N_C} + 2n_pt_{oh}.
\end{equation}

\begin{figure}[t]
    \centering
    \includegraphics[width=8.5cm]{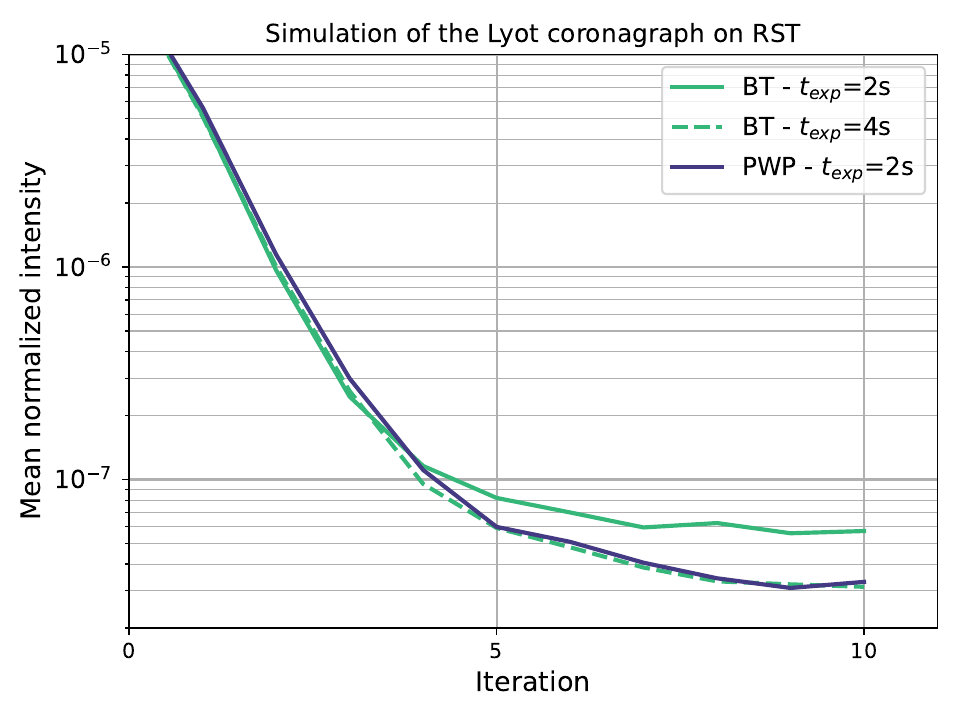}
    \caption{Simulated correction of the Roman Space Telescope Lyot Coronagraph diffraction pattern through PWP and BTP. We assume a 7-mag star is observed in narrowband (1\%) at 575~nm. Detector noise is considered null. $t_{exp}$ represents the exposure time of each acquired probed image. Since BTP needs half the number of probe images used in PWP, $t_{cal}$ is therefore equal for both BTP and PWP.}
    \label{fig:ContrastvsIterSimu}
\end{figure}

Comparing both eq.~\ref{eq:timeBTP} and eq.~\ref{eq:timePWP} shows that, for negligible overheads, PWP is as efficient as BTP to obtain the desired contrast level even though it demands a much higher number of images. The exposure time for each probe image needs to be twice longer in the case of BTP with respect to PWP to reach similar contrast. This calculation is confirmed with a FALCO\cite{Riggs2018} simulation whose results are presented in Fig.~\ref{fig:ContrastvsIterSimu} in the context of RST. Here, two 48x48 actuator DMs in cascade aim for correcting the RST diffraction pattern in a 360$\degree$ dark hole ranging from 2.7 to 10~$\lambda/D$. We consider a Lyot coronagraph whose occulting mask diameter is 5.4~$\lambda/D$. We assume photon noise only while observing a $M_V=7$ star in 1\% bandwidth at 575~nm. This simulation verifies that each individual exposure time needs to be doubled for BTP with respect to PWP to retrieve similar contrast levels, leading to an identical total exposure time for both the algorithms. However, in cases where overheads time dominate i.e. the algorithm efficiency is not limited by photon noise (for instance in the fictional scenario where RST faces long data transfer from the spacecraft to the ground), BTP seems to be a promising technique to create dark hole regions in a smaller amount of time.

\subsubsection{If \texorpdfstring{$| iC\{A\Psi\} |^2$}{Lg} is measured at first iteration only}
Finally, we assume an in between case where $| iC\{A\Psi_p\} |^2$ is calibrated once during the first iteration following eq.~\ref{eq:modelfreeprobe}. It means $I^-$ would be measured only once for the whole BTP+EFC sequence. Then, according to eq.~\ref{eq:deltaI_glob}:
\begin{equation}
    \sigma^2\{\Delta I\} = \sigma^2\{2(I^+-I^0-\frac{I^+(0)-I^-(0)}{2}+I^0(0))\}=\sigma^2\{2I^+ - I^+(0) - I^-(0)\}=6p^2
\end{equation}
where $I^+(0)$, $I^-(0)$ and $I^0(0)$ stand for the positive, negative and unprobed images at first iteration. Then, according to eq.~\ref{eq:totalvariance2}:
\begin{equation}
    \sigma^2\{E_R\} = \frac{3}{4n_p}
\end{equation}
Again, we can now compute eq.~\ref{eq:NC}:
\begin{equation}
    N_C = \frac{3}{2n_pF_{pk}t_{exp}} = \frac{3}{2F_{pk}t_{cal}}
\end{equation}
to compute the total exposure time required per iteration:
\begin{equation}
    t_{tot} = t_{cal} + t_{toh} = \frac{3}{2F_{pk}N_C} + n_pt_{oh}.
\end{equation}
That case becomes interesting if one wants to mitigate the impact of model errors while using BTP because of large overheads.

\subsection{Sensitivity to model error}
\label{subsec:modelerror}
We first assumed that model errors were negligible. We now take into account their repercussion on the E-field error by combining eq.~\ref{eq:totalvariance} and eq.~\ref{eq:varianceBTP}, when $| iC\{A\Psi\} |^2$ is computed from the propagation model. The variance of the estimated E-field in the BTP case is therefore:
\begin{equation}
    \begin{pmatrix}
\sigma^2\{E_R\} \\
\sigma^2\{E_I\}
\end{pmatrix} =  \left(\frac{1}{2n_p}\right) \left(1+4\sigma^2\{p\}\right)\begin{pmatrix}
1 \\
1
\end{pmatrix}+\frac{\sigma^2\{p\}}{p^2} \begin{pmatrix}
E_R^2 \\
E_I^2
\end{pmatrix}
\end{equation}
Let also roughly assume that the model error on the probe amplitude is proportional to the probe amplitude ($\sigma^2\{p\} = \alpha p^2$), i.e. there is a small gain misregistration. In that case, we can write:
\begin{equation}
    \begin{pmatrix}
\sigma^2\{E_R\} \\
\sigma^2\{E_I\}
\end{pmatrix} =  \left(\frac{1}{2n_p}\right) (1+\alpha p^2)\begin{pmatrix}
1 \\
1
\end{pmatrix}+\alpha \begin{pmatrix}
E_R^2 \\
E_I^2
\end{pmatrix}.
\end{equation}
On the other hand, in the case of PWP:
\begin{equation}
    \begin{pmatrix}
\sigma^2\{E_R\} \\
\sigma^2\{E_I\}
\end{pmatrix} =  \left(\frac{1}{4n_p}\right)\begin{pmatrix}
1 \\
1
\end{pmatrix}+\alpha \begin{pmatrix}
E_R^2 \\
E_I^2
\end{pmatrix}.
\end{equation}

Again, we can compare the two last equations. While the error on the E-field estimate does not depend on the probe amplitude in the PWP case, we notice the error in the BTP estimates increases with the probe intensity. Therefore, depending on the level of model errors, the probe intensity (or amplitude) might need to be diminished with respect to PWP to get an accurate E-field estimate. This would induce longer exposure time to compensate, hence a smaller efficiency with respect to PWP.

\subsection{Implicit method}
\label{subsec:implicit}
Mitigating the impact of model errors can be performed through a data-driven (or implicit) BTP+EFC, as previously derived for PWP+EFC\cite{Ruffio2022,Haffert2023}. We recall eq.~\ref{eq:inversion} that links the $\Delta I$ of eq.~\ref{eq:deltaI_glob} to the speckle E-field:
\begin{equation}
    z = Hx.
\end{equation}
The speckle E-field can also be decomposed on a basis of modes created on the DM: $x = G a$, where $a$ are the DM commands. Therefore, we can write:
\begin{equation}
    z = HGa = Ka,
\end{equation}
where $K = HG = \frac{\partial z}{\partial a}$ represents the Jacobian that transforms the command applied on the DM to the $\Delta I$ in eq.~\ref{eq:deltaI_glob} and which can directly be calibrated through measurements. Measuring this response matrix is performed by applying each mode $m$ individually through the following recipe: 
\begin{equation}
    \Delta(z_k)_m = \frac{1}{2b}\left[\left(I_{+m}^{+k} - I_{+m}^0 - | iC\{A\Psi_k\} |^2\right) -  \left(I_{-m}^{+k} - I_{-m}^0 - | iC\{A\Psi_k\} |^2\right) \right]
\end{equation}
where $I_{\pm m}^{+k}$ means the acquired science image while applying the $k^{th}$ probe and the $m^{th}$ mode (either positive or negative). $I_{\pm m}^0$ means that the probe is not applied on top of the mode $m$ before taking the image. $b$ stands for the modal voltage that is being applied. We notice this vector basis can be simplified in:
\begin{equation}
    \Delta(z_k)_m = \frac{1}{2b}\left[\left(I_{+m}^{+k} - I_{-m}^{+k}\right) -  \left(I_{+m}^0 - I_{-m}^0\right) \right]
\end{equation}
requiring 4 images per probe per mode. But, we also notice that no probe is applied in $I_{+m}^0$ and $I_{-m}^0$. Therefore, they can be computed once to work for every probe saving $2n_p-2$ images per mode with respect to iEFC while calibrating the Jacobian. Again, this aspect is particularly interesting when the time to perform such calibration is dominated by overheads.

\section{Experimental tests in stabilized environments}
\label{sec:JPLexperiment}
\subsection{Vortex coronagraphs on the In-Air Coronagraph Testbed}
\label{subsec:Vortex}
RST HOWS\&C will be performed via commands for the ground and its software has been designed to allow testing of various types of control approaches\cite{Kasdin2020}. To understand if BTP is a viable option for space-based applications, we aim to compare BTP's behaviour in stabilized environment at contrast levels of $\sim10^{-8}$ and below. We therefore tested BTP in diverse conditions and for various coronagraphs at the High-Contrast Imaging Testbed facility located at JPL.

\begin{figure}[t]
    \centering
    \includegraphics[width=8cm]{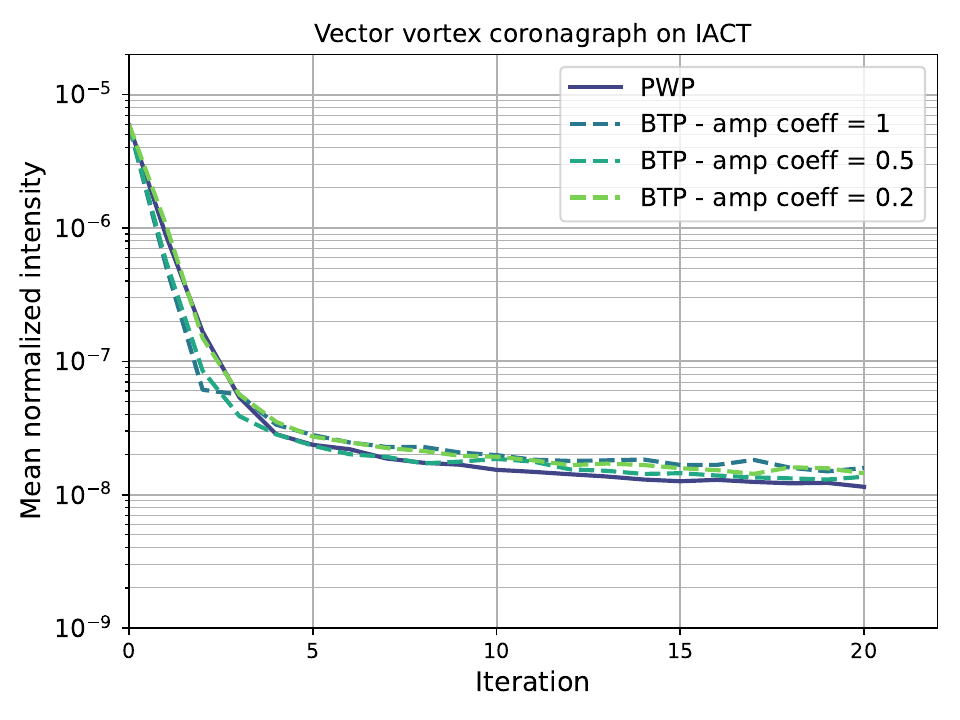}
    \includegraphics[width=8cm]{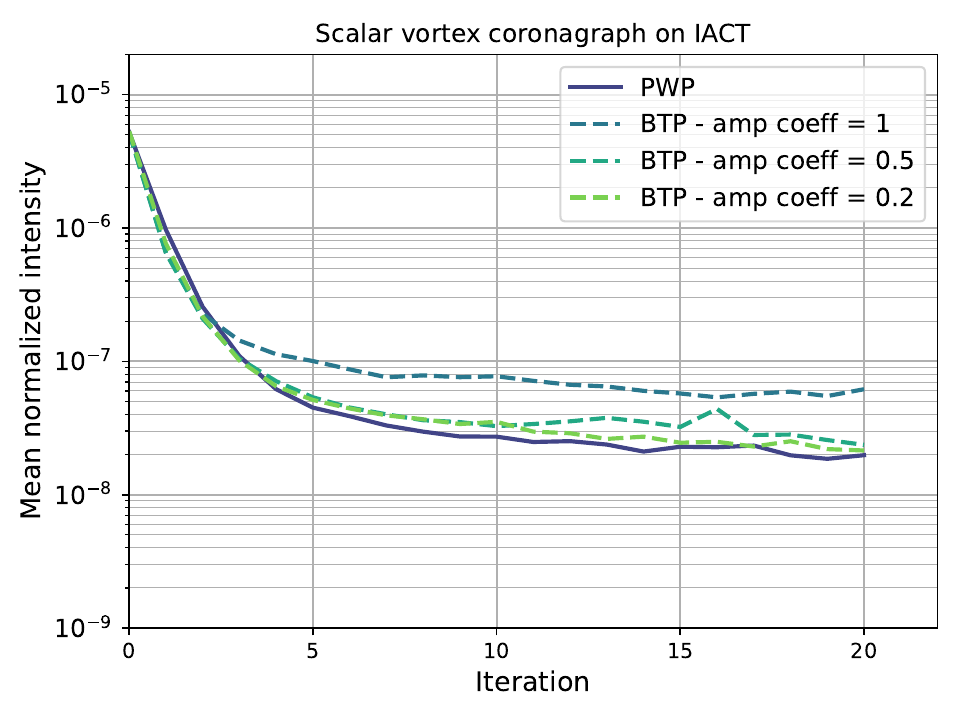}
    \caption{Mean normalized intensity in the image region corrected by BTP+EFC and PWP+EFC on IACT. A vector (left) and scalar (right) coronagraphs are used. "amp coeff" is the factor applied to the probe amplitude that PWP would have been used at each iteration.}
    \label{fig:ContrastvsIterIACT}
\end{figure}

Using FALCO, we have tested BTP in association with EFC with a charge 6 Vector Vortex Coronagraph\cite{Mawet2009} (VVC) on the In-Air Coronagraph Testbed\cite{Baxter2021} in a half dark hole configuration. The experiment is performed in narrow band (1\%) around 635~nm. The DH ran from 4 to 15$\lambda/D$ because a closer edge was not converging for BTP, unlike PWP with equivalent probe amplitude. BTP has been configured model-based (see in Sec.~\ref{subsubsec:modelbased}) while the model-free flavours of BTP presented above remain to be tested. For both algorithms, the probe amplitude is set proportional to the current contrast and therefore decreases along the iterations\cite{Groff2015}. Identical probe shapes (sinc functions) are also being used. Figure~\ref{fig:ContrastvsIterIACT} (left) shows the convergence of the algorithm for various probe amplitudes as well as for PWP. No significant difference can be derived from those curves. We therefore can assume the VVC6 propagation model is of high fidelity. 

Contrariwise, same experiment has been performed with a charge 6 Scalar Vortex Coronagraph\cite{Errmann2013} in narrow band (1\%) around 775~nm (see in Fig.~\ref{fig:ContrastvsIterIACT}, right). For the SVC, a 80\% decrease of the probe amplitude has been needed to reach similar performance than PWP. Indeed, we know the manufacturing of SVC introduces defects in the mask that are not yet understood in the models\cite{Desai2023b}. If PWP seems robust to this discrepancies, BTP's performances further suffer from these model errors as explained in Sec.~\ref{subsec:modelerror}. Either the implicit BTC (see in Sec.~\ref{subsec:implicit}) or measuring $| iC\{A\Psi_k\} |^2$ at the first iteration (via eq.~\ref{eq:modelfreeprobe}) could mitigate that issue.

\begin{figure}[t]
    \centering
    \includegraphics[width=8cm]{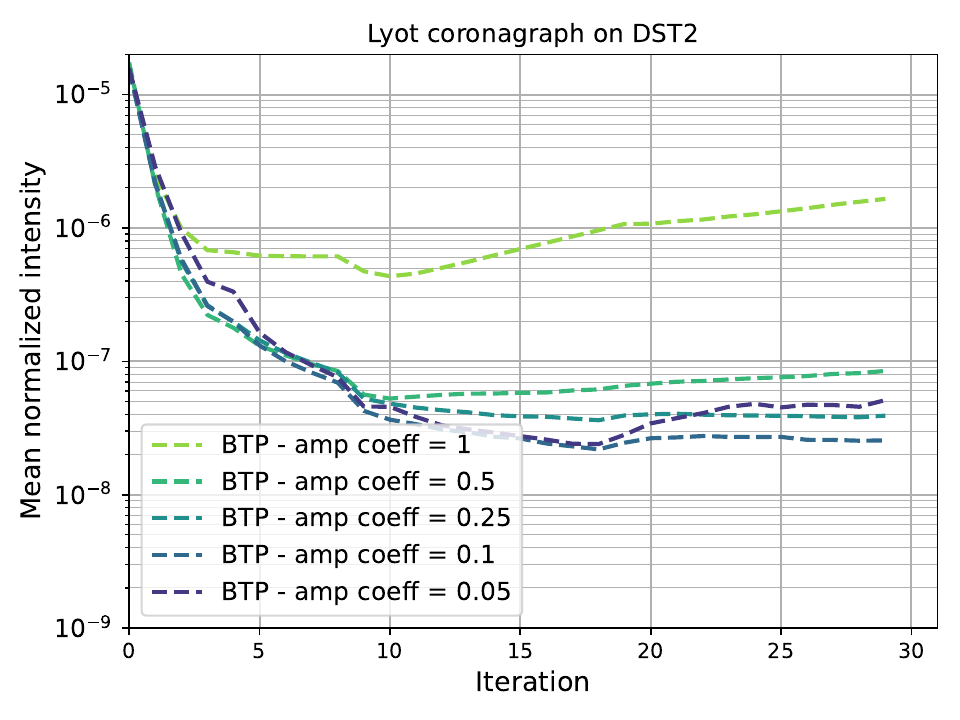}
    \includegraphics[width=8cm]{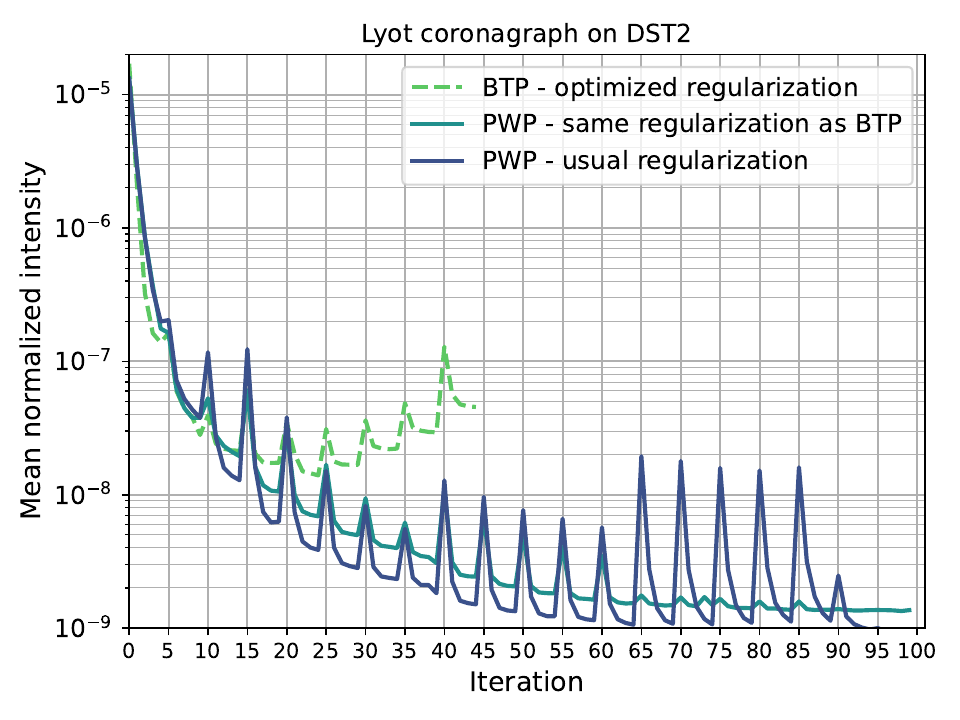}
    \caption{Mean normalized intensity in the image region corrected by BTP+EFC and PWP+EFC on DST2 equipped with a Lyot coronagraph. On the left, no $\beta$-bumping regularization is being applied. "amp coeff" is the factor applied to the probe amplitude that PWP would have been used at each iteration. On the right, two different $\beta$-bumping regularizations are used: the first $\beta$-bump is topical for PWP while the second presents a weaker relaxation for BTP to converge.}
    \label{fig:ContrastvsIterLyot}
\end{figure}

\subsection{Lyot coronagraph on the Decadal Survey Testbed 2}
\label{subsec:LC}
An equivalent experiment has been conducted under vacuum on the Decadal Survey Testbed 2\cite{Meeker2021, Noyes2023} (DST2) with a Lyot coronagraph whose radius is 2.7$\lambda/D$ at 660nm. The correction region is also a half dark hole running from 2.7 to 15 $\lambda/D$. Again, algorithm convergence is presented in Fig.~\ref{fig:ContrastvsIterLyot} (left) for decreasing probe amplitudes. In the Lyot coronagraph case, taking PWP-like probe amplitudes cause the BTP+EFC loop to diverge after 10 iterations. Probes whose amplitudes are ten times lower than in the PWP case are required for the contrast to reach $\sim2\cdot10^{-8}$. Such contrast is however one order of magnitude worse than the best contrast obtained with PWP. DST2 is indeed able to reach contrast down to $10^{-9}$ with the Lyot coronagraph (see in Fig.~\ref{fig:ContrastvsIterLyot}, right) taking advantage of the $\beta$-bumping technique where the regularization is regularly relaxed to correct for higher order modes that would otherwise limit the contrast performance\cite{Sidick2017b}. We therefore also tried an equivalent regularization scheme with BTP but the loop quickly diverged. We eventually managed to combine BTP with $\beta$-bumping through a much weaker relaxation to reach contrast levels of $\sim 1\cdot10^{-8}$ even though the loop quickly diverged again after obtaining this contrast record.

Full model-based BTP (see in Sec.~\ref{subsubsec:modelbased}) is therefore currently not suitable to reach contrast levels relevant for space-based applications because it is not robust enough to model errors. Implicit or calibrated methods presented earlier remain to be studied in such stabilized environments.

\begin{figure}[t]
    \centering
    \includegraphics[width=\linewidth]{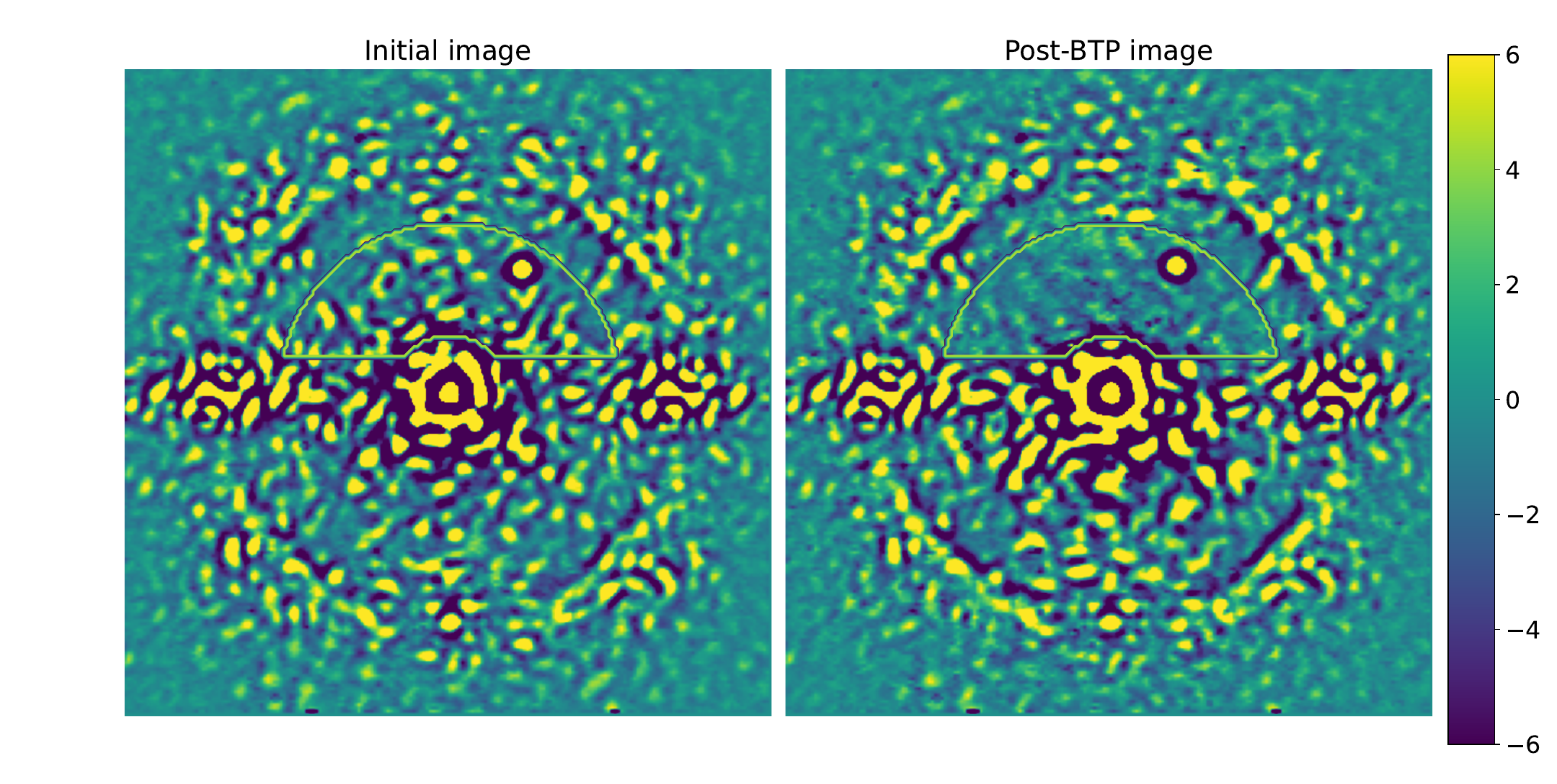}
    \caption{($\times10^{-6}$). High-pass filtered normalized image of $\beta$-pictoris observed with VLT/SPHERE in its H3/APLC configuration before (left) and after (right) correction of the static speckles with BTP. The region where the correction were performed is encircled in bright green. The bright dot remaining in the post-BTP dark hole region corresponds to $\beta$-pictoris b.}
    \label{fig:BTP}
\end{figure}

\section{Experimental tests under atmospheric turbulence}
\label{sec:SPHEREexperiment}
Section~\ref{subsec:sensitivity} has shown that BTP is more efficient than PWP if the time required to take a science image is dominated by overheads. An interesting case is therefore the on-sky compensation of non-common-path aberrations for high-contrast imaging applications. Indeed, the atmospheric turbulence must be frozen through long exposure times to reveal the static and quasi-static speckles that are targeted by the FPWFS \cite{Singh2019,Potier2019,Potier2022b}~. Additionally, contrast performance below $10^{-8}$ are anyway currently unattainable with ground-based instruments due to the atmospheric turbulence. These arguments makes BTP a promising FPWFS for such facilities.

We therefore tested BTP on VLT/SPHERE\cite{Beuzit2019} while observing $\beta$-Pictoris on March 31st 2023. Observing conditions where rather poor, forcing the operation of the medium pinhole for spatial filtering in front on the Shack-Hartmann. We used the Apodized Pupil Lyot coronagraph whose focal plane occultor is 185mas in diameter, in the H3 band ($\lambda_0 = 1667$nm, $\Delta\lambda= 55$nm). BTP+EFC is performed in a half dark hole whose inner (resp. outer) edge is 220mas (resp. 650mas), encompassing the known position of $\beta$-Pictoris b. Two probes (poking of neighbor actuators) were employed with a peak-to-valley amplitude of 400~nm, as previously used for PWP\cite{Potier2022b}. The exposure time for each image is 32s while one iteration (acquisition of three images + overheads) takes a total of 140s. We also inform the reader that pixel windowing was applied to decrease the detector readout time with respect to previous work\cite{Potier2022b}~.

\begin{figure}[t]
    \centering
    \includegraphics[width=12cm]{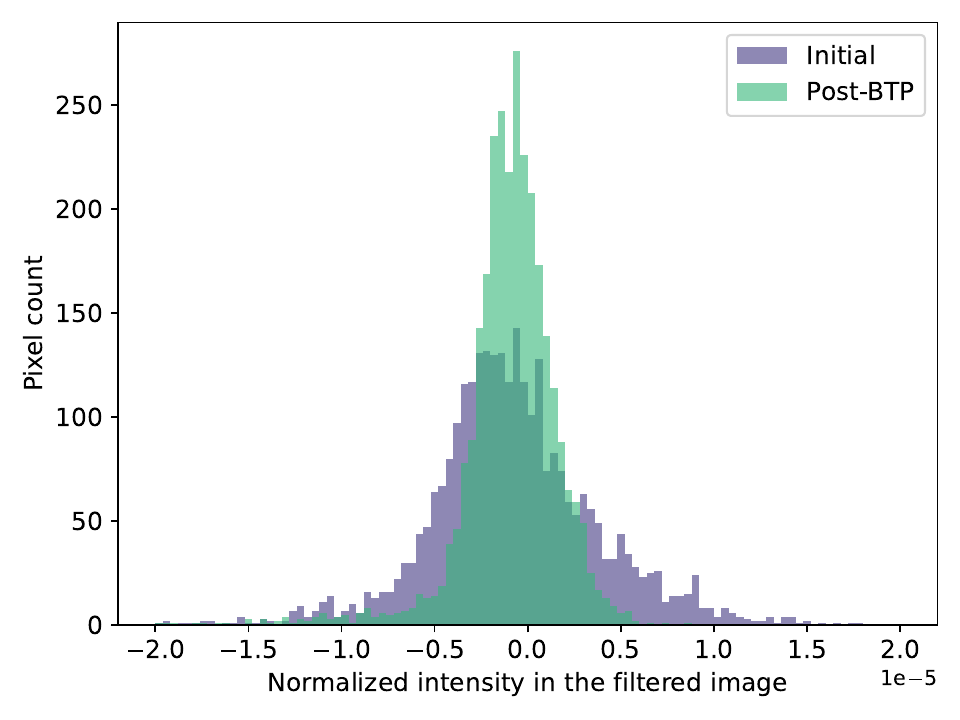}
    \caption{Statistics of the pixels in the dark hole region before (purple) and after (green) correction of the static speckles with BTP. Pixel value is in normalized intensity.}
    \label{fig:BTP_histo}
\end{figure}

Figure~\ref{fig:BTP} shows the result of BTP+EFC after 4 iterations ($\sim10$ minutes) where the images have been high-pass filtered to emphasize the correction of static speckles. Qualitatively, BTC+EFC attenuated an important part of the speckle intensity in the Half Dark Hole, increasing the signal to noise for $\beta$-Pictoris b detection (whose image remains unaffected by the algorithm). Pixel statistic is presented in Fig.~\ref{fig:BTP_histo} in the form of an histogram where pixels at planet position are removed. The standard deviation has decreased from $4.70\cdot10^{-6}$ to $2.97\cdot10^{-6}$ representing a 1.6 improvement ratio. After these 4 iterations, the loop started to diverge. In this configuration, it seems BTP then presented weaker performance than PWP in previous experiments\cite{Potier2022b}. But it is also worth noting that atmospheric conditions where far from ideal during this experiment. In the future, smaller probe amplitude will be employed to make the algorithm more robust to model errors in poor observing conditions. Also, since the probe shapes and amplitudes are constant during the DH calibrations, the data-driven calibration of the probe intensity could also be employed at first iteration (see eq.~\ref{eq:modelfreeprobe}).

\section{Conclusion}
\label{sec:conclusion}
In this work, we revisited the Borde \& Traub (BTP) focal plane wavefront estimation technique for the sake of high-contrast imaging. After deriving equivalence of BTP and pair wise probing (PWP) with respect to photon noise sensitivity, we show that BTP is in theory more efficient than PWP in the presence of image overheads. However, BTP is also more sensitive to model errors as demonstrated during the various dark hole digging experiments at JPL/HCIT. This downside might prevent its use when high contrast ratio are required, for instance when looking for temperate exoplanets. However, the BTP model-free flavors could represent promising alternatives toward high-contrast ratios.

Nevertheless, model-based BTP has demonstrated $\sim10^{-8}$ contrast levels on the different testbeds, which makes it a hopeful option to attenuate the quantity of static aberrations in ground-based instruments. Indeed the correction of non common path aberration through FPWS\&C might drastically limit the instrument duty cycle since 1)the image exposure time is increased artificially to average out and suppress the dynamic speckles caused by the atmospheric turbulence and 2)slow science detectors are mainly used. Because BTP requires less images in total, such algorithm could outperform PWP if the propagation model is accurate and small enough probes are used.

\acknowledgments 
The research was carried out at the Jet Propulsion Laboratory, California Institute of Technology, under a contract with the National Aeronautics and Space Administration (80NM0018D0004). This work has been carried out within the framework of the NCCR PlanetS supported by the Swiss National Science Foundation under grants 51NF40\_182901 and 51NF40\_205606.


\bibliography{bib_GS}   
\bibliographystyle{spiejour}   



\end{document}